\documentclass[twocolumn,epsfig,superscriptaddress]{revtex4}

\usepackage{graphicx}
\usepackage{amssymb}
\usepackage{amsmath}
\usepackage{color}
\usepackage{bm}
\usepackage{physics}
\usepackage{changes}

\newcommand{\figsizeone}{0.48}
\newcommand{\figsizethree}{0.8}

\makeatletter
\let\Changes@Markup@Deleted\@gobble
\makeatother

\begin{document}

\draft
\title{Non-orientability induced PT phase transition in ladder lattices}
\author{Jung-Wan Ryu}
\affiliation{Center for Theoretical Physics of Complex Systems, Institute for Basic Science, Daejeon 34051, Republic of Korea}
\author{Nojoon Myoung}
\affiliation{Department of Physics Education, Chosun University, Gwangju 61452, Republic of Korea}
\author{Martina Hentschel}
\affiliation{Institute of Physics, Faculty of Natural Sciences, 
Technische Universit\"at Chemnitz, Reichenhainer Str. 70, 09126 
Chemnitz, Germany}
\author{Hee Chul Park}
\email{hcpark@ibs.re.kr}
\affiliation{Center for Theoretical Physics of Complex Systems, Institute for Basic Science, Daejeon 34051, Republic of Korea}
\date{\today}

\begin{abstract}

We study parity-time (PT) phase transitions in the energy spectra of ladder lattices caused by the interplay between non-orientability and non-Hermitian PT symmetry. 
The energy spectra show level crossings in circular ladder lattices with increasing on-site energy gain-loss because of the orientability of a normal strip. 
However, the energy levels show PT phase transitions in PT-symmetric M{\"o}bius ladder lattices due to the non-orientability of a M{\"o}bius strip. 
In order to understand the level crossings of PT symmetric phases, we generalize the rotational transformation using a complex rotation angle.
We also study the modification of resonant tunneling induced by a sharply twisted interface in PT-symmetric ladder lattices.
Finally, we find that the perfect transmissions at the zero energy are recovered at the exceptional points of the PT-symmetric system due to the self-orthogonal states.
\end{abstract}
\maketitle
\narrowtext

\section{Introduction}

Non-Hermitian physics has attracted considerable attention not only as an important alternative to the Hermitian formalism of quantum mechanics in open systems with energy gain and loss \cite{Moi11} but also for applications in various optical and photonic systems \cite{Guo09, Rue10, Lie12, Bra14, Hod17, Che17}. In a non-Hermitian system, the eigenvalues are complex and the eigenstates form a biorthogonal set. The complex eigenvalues have a clear physical meaning: the real and imaginary parts represent the eigenenergy of a state and its decay rate, respectively.

The M{\"o}bius strip, discovered independently by M{\"o}bius and Listing in 1858, is a continuous one-sided surface formed by rotating one end of a rectangular strip through 180 degrees and attaching it to the other end \cite{Moe65, Lis61}.
This marvelous structure with only one side and only one boundary is the epitome of a topologically non-trivial system and shows curious properties due to its non-orientability. A surface in Euclidean space is orientable if a two-dimensional figure cannot be moved around the surface and back to where it started so that it looks like its own mirror image \cite{Spi65, Hat01}. Otherwise, the surface is non-orientable. Besides fundamental studies on topology, the non-orientability of M{\"o}bius strips has enabled their use in many applications in various fields \cite{Pic06, Wal82, Gil15, Han10}.

The real-space M{\"o}bius strip is therefore a fascinating system both for mathematicians and physicists. Physics in particular allows one to realize M{\"o}bius strip topologies in various contexts beyond real space. For example, an exceptional point (EP), which is a degenerate point of eigenenergies in a non-Hermitian system, generates this structure in parametric space because of the square root branch property of the singular point \cite{Kat96, Hei12}. Such a M{\"o}bius strip of eigenenergies generated in parameter space has been reported in microwave experiments \cite{Dem01, Dem03}, optical microcavities \cite{Lee09}, and a chaotic exciton-polariton billiard \cite{Gao15}. One of the interesting phenomena originating from the non-trivial topology of non-Hermitian systems is parity-time (PT) symmetry, which exhibits a spontaneous symmetry-breaking transition from an unbroken PT symmetric phase to a broken phase via EPs \cite{Ben98}. PT symmetry is protected in non-Hermitian systems with a balance of energy gain and loss represented by the commutation relation [H, PT] = 0, where H is a Hamiltonian. Many PT-symmetric systems have been explored in several fields, including optics \cite{Elg07, Mus08, Mak08, Kla08, Guo09, Rue10, Reg12}, electronic circuits \cite{Sch11}, atomic physics \cite{Jog10}, and magnetic metamaterials \cite{Laz13}.

While the non-trivial M{\"o}bius topology in real space has been widely applied in chemistry \cite{Wal82, Gil15} and biology \cite{Han10}, there have been only a few related studies in physics \cite{Zha09, Guo09b, Kre18}. Most of these studies were particularly focused on the topological properties of quantum states in Hermitian systems \cite{Zha09, Guo09b}. In this paper, we study how the non-orientability of a M{\"o}bius strip generates and affects the PT phase transitions in tight-binding models by comparing circular and M{\"o}bius ladder lattices as the simplest models in real space. Finally, corresponding resonances and antiresonances appearing in quantum transport in quasi-one-dimensional lattices are also studied and compared to the energy spectra in circular and M{\"o}bius ladder lattices.

This paper is organized as follows. In Section II, we introduce the two systems, circular and M{\"o}bius ladder lattices, and then review their characteristics. In Section III, we study eigenenergy evolution as a function of on-site potential in PT-symmetric cases, and in Section IV we consider quantum transport in PT-symmetric ladder and twisted-ladder lattices corresponding to PT-symmetric circular and M{\"o}bius ladder lattices, respectively. Section V summarizes our results.

\begin{figure}
\begin{center}
\includegraphics[width=\figsizeone\textwidth]{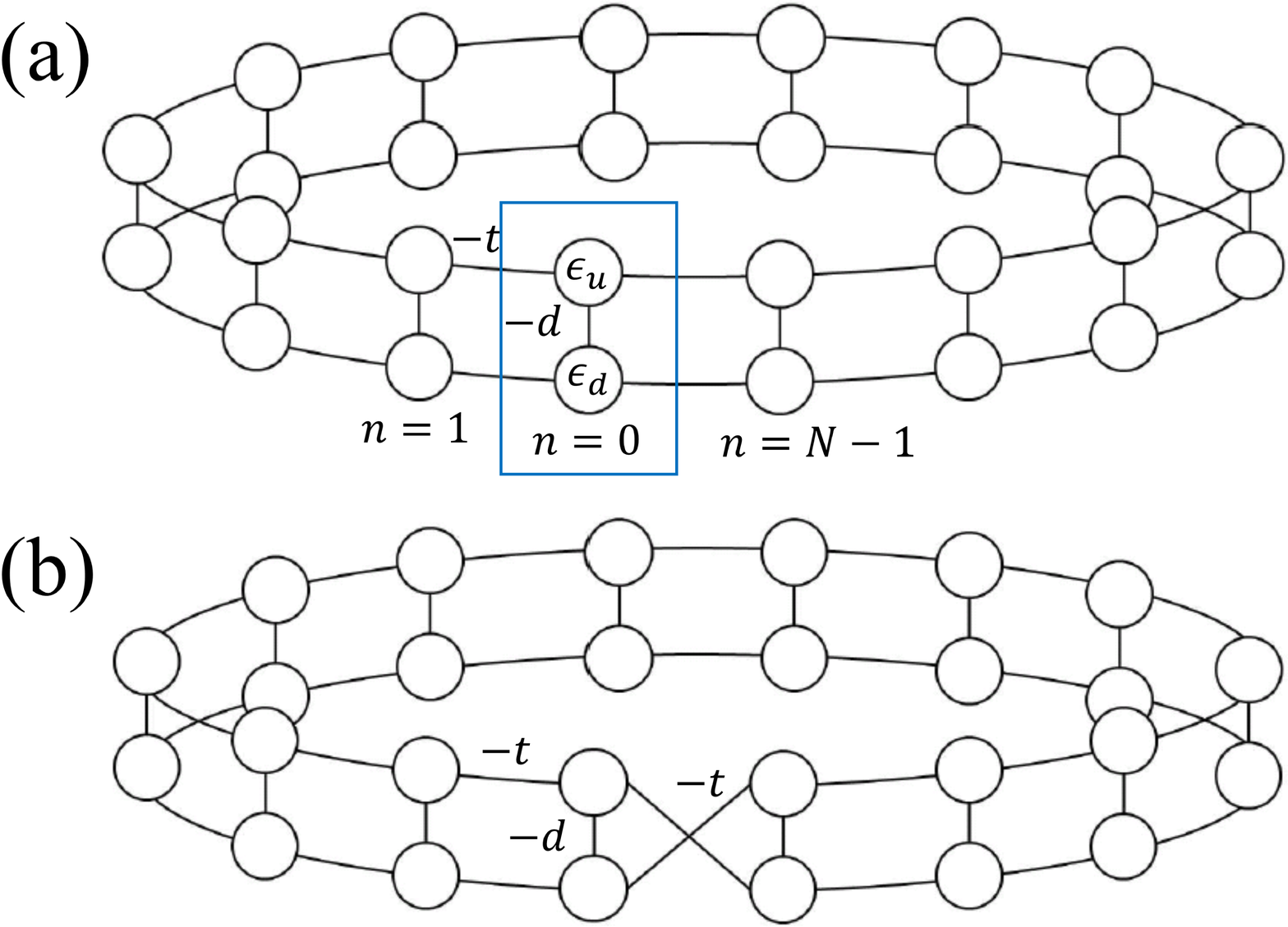}
\caption{(color online).
(a) A circular ladder lattice (CLL) and (b) a M{\"o}bius ladder lattice (MLL). The blue box represents the unit cell of the CLL, which has two sites with intra-unit cell hopping strength $d$ and inter-unit cell hopping strength $t$.
}
\label{fig1}
\end{center}
\end{figure}

\section{systems: circular and M{\"o}bius ladder lattecs}

Figure~\ref{fig1} (a) depicts a circular ladder lattice (CLL). The on-site potentials of the upper and lower sites are $\epsilon_u$ and $\epsilon_d$, respectively, while the intra- and inter-unit cell hopping strengths are $d$ and $t$, respectively. The Hamiltonian of an infinite ladder lattice can be expressed on the basis of Pauli matrix $\bm{\sigma} = (\sigma_x, \sigma_z)$ and $k$-independent vector field $\bm{h}=(-d, \delta/2 + i \gamma/2)$ as
\begin{equation}
 H(k)=\bm{h} \cdot \bm{\sigma} + h_0 (k) \sigma_0,
\label{Ham}
\end{equation}
where the extra term $h_0(k) = -t \cos{k}$ and $\sigma_0$ is an identity matrix. $\delta$ and $\gamma$ will be used to describe antisymmetric on-site potentials when we consider Hermitian and PT-symmetric situations. The eigenvalues of the Hamiltonian are
\begin{equation}
 \varepsilon_\pm = \pm |\bm{h}| + h_0 (k) = -2 t \cos{k} \pm \sqrt{d^2 + \left(\frac{\delta}{2}+i \frac{\gamma}{2}\right)^2},
 \label{ev}
\end{equation}
which are complex with $\varepsilon=\varepsilon_r + i \varepsilon_i$. If $d=0$, a CLL can be divided into two circular lattices with hopping strength $t$. 

Next we consider the M{\"o}bius ladder lattice (MLL) shown in Fig.~\ref{fig1} (b) \cite{Guy67}. On-site potentials and inter- and intra-unit cell hopping strengths are the same as those in the CLL; the only difference is that one pair of parallel hoppings changes into cross hoppings. In this paper, we model the M{\"o}bius strip with an abrupt change in the hopping parameters and leave the study of a smooth parameter change to a subsequent work. If $d=0$, the MLL can be considered as one circular lattice with two times the length of the MLL, which reflects the intrinsic properties of a M{\"o}bius strip.

We consider a CLL with symmetric on-site potential, i.e., $\epsilon_u=\epsilon_d=0$. The energy bands of the CLL are separated into two sub-bands in which states are equally distributed to upper and lower lattices with odd and even parities, respectively. As $d$ and $t$ increase, the distance between the two bands and the band widths increase, respectively. Using a periodic boundary condition, the eigenvalues of the Hamiltonian of a CLL with $N$ sites are
\begin{equation}
 \varepsilon_\pm = -2 t \cos{\frac{2 n \pi}{N}} \pm d \quad\quad (n= 1, \cdots, N),
 \label{ev_cll}
\end{equation}
where $\varepsilon_{\pm}$ are the eigenvalues of the corresponding eigenstates with odd and even parities, i.e., $u_u (x) = - u_d (x)$ and $u_u (x) = u_d (x)$, respectively.

Eigenenergies in the MLL can be obtained by replacing the boundary condition for a CLL with that for an MLL. As a result, the eigenvalues of the Hamiltonian of an MLL with $N$ sites are
\begin{eqnarray}
 \label{ev_mll}
 \nonumber
 \varepsilon_- &=& -2 t \cos{\frac{2 n \pi}{N}} - d \quad\quad\quad \text{for even parity,} \\
 \varepsilon_+ &=& -2 t \cos{\frac{(2n-1) \pi}{N}} + d \quad \text{for odd parity,}
\end{eqnarray}
where $n= 1, \cdots, N$. The Bloch wave vectors and corresponding eigenenergies are the same in the case of even-parity eigenstates but different in the case of odd-parity eigenstates. The upper and lower sites exhibit opposite signs of amplitudes in the case of odd-parity eigenstates, which correspond to opposite orientations, while the sites have the same signs of amplitudes (meaning no orientations) in the case of even-parity eigenstates. Consequently, for the even-parity case, there is no effect of breaking orientation in the MLL since there is no orientation even in the CLL. This changes when odd-parity states are considered. It is noted that the symmetry of the unit cell plays an important role in obtaining eigenenergies.

The difference between normal rings and M{\"o}bius ring structures according to the parity of the eigenstates has previously been studied \cite{zli12}. It has also been found that the difference in eigenenergies in the case of odd-parity in a M{\"o}bius ring structure is related to the non-integer azimuthal mode indices in three-dimensional optical M{\"o}bius strip cavities \cite{Kre18}. In this case, the polarization of light induces an orientation, such as opposite signs of the amplitudes of the odd-parity eigenstates. In the next section, we study the PT phase transitions of complex eigenenergies in PT-symmetric cases.

\section{Energy spectra in PT-symmetric circular and M{\"o}bius ladder lattices}

\begin{figure}
\begin{center}
\includegraphics[width=\figsizeone\textwidth]{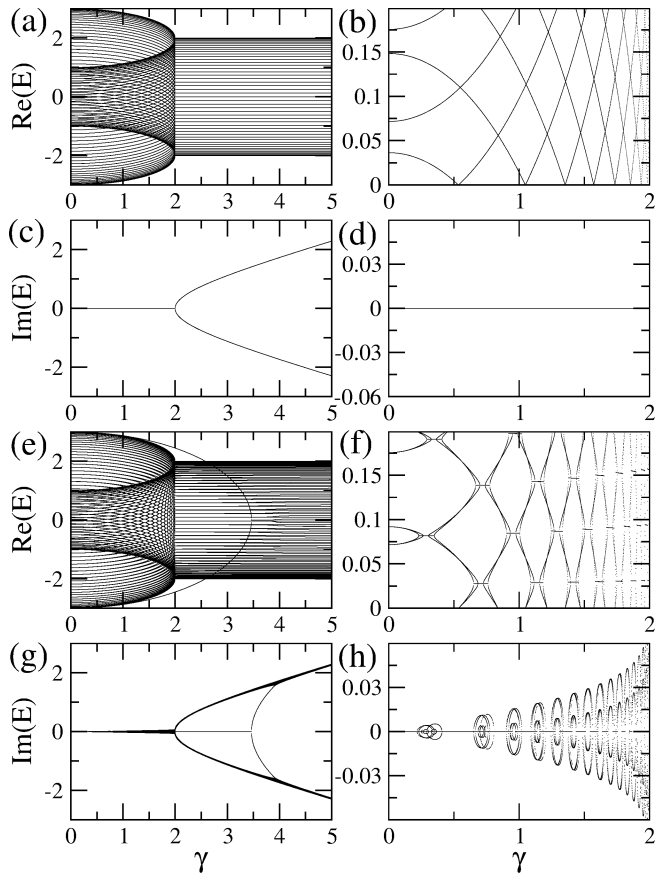}
\caption{(a) Real and (c) imaginary parts of the eigenenergies as a function of $\gamma$ in a PT-symmetric CLL with $100$ unit cells when $d=t=1$. (b) Real and (d) imaginary parts of the selected eigenenergies. (e) Real and (g) imaginary parts of the eigenenergies as a function of $\gamma$ in a PT-symmetric MLL with $100$ unit cells when $d=t=1$. (f) Real and (h) imaginary parts of the selected eigenenergies. See also Fig.~\ref{fig3} and \ref{fig4} below.}
\label{fig2}
\end{center}
\end{figure}

\begin{figure}
\begin{center}
\includegraphics[width=\figsizeone\textwidth]{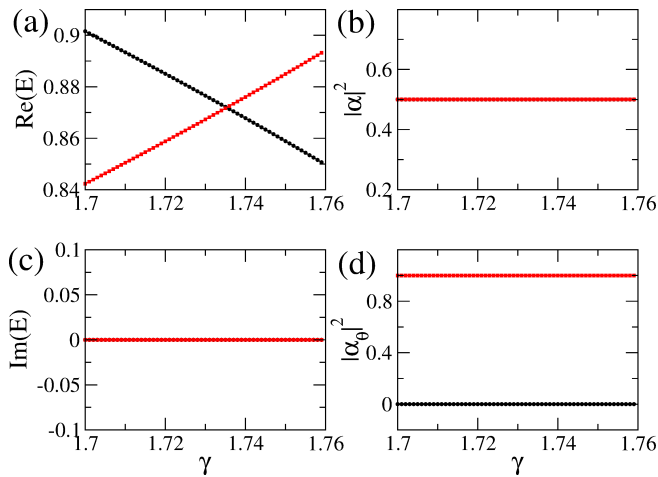}
\caption{(color online).
(a) Real and (c) imaginary parts of the eigenenergies in a PT-symmetric CLL showing level crossings of real eigenenergies of a non-Hermitian system in the PT-unbroken phase. (b) $|\alpha|^2$ of four degenerate eigenstates and (d) $|\alpha_\theta|^2$ of four degenerate eigenstates.}
\label{fig3}
\end{center}
\end{figure}

We consider PT-symmetric systems that can possess entirely real eigenvalues, which is a crucial factor in the transport problems of the next section. A PT-symmetric CLL with balanced gain and loss is introduced with antisymmetric imaginary on-site potentials in Eq.~(\ref{Ham}), i.e., $\epsilon_u = -\epsilon_d = i \gamma /2$. Here, the systems parameters are $d=t=1$. A PT-symmetric CLL exhibits complex energy bands because they are non-Hermitian systems. Figure~\ref{fig2} (a) and (c) show a PT phase transition between unbroken and broken PT-symmetric phases possessing real and complex conjugate eigenenergies, respectively. As $\gamma$ increases in the range of $0<\gamma < \gamma_b$, pairs of real energies approach each other to ultimately merge at a corresponding PT phase transition point, $\gamma_b = 2d$. After the transition at $\gamma = \gamma_b$, degenerate energies split into two complex conjugate energies, and the absolute values of the imaginary parts increase as $\gamma$ increases. The two bands are separable, and thus there is no band touching except for the case of the EPs, $\gamma = \gamma_b$, in the PT-symmetric CLL. In the case of the unbroken phase, there is real energy band separation because the imaginary parts of the energy bands equal to zero, and thus their states are equally distributed to the upper and lower lattices in spite of the PT-symmetric on-site potential [see Fig.~\ref{fig3} (b)]. In the case of the broken phase, there is imaginary energy band separation because the real parts of the energy bands are the same. The eigenstates of the positive and negative imaginary energy bands have gain and loss, respectively, and thus their states are not equally distributed to the upper and lower circular lattices. It is noted that PT-symmetric on-site potential in a PT-symmetric CLL makes the real parts of the energies attractive, as shown in Fig.~\ref{fig2} (a), while antisymmetric real on-site potential in a CLL makes the real energies repulsive \cite{Ryu17a}.

Considering only the unbroken region of $0 < \gamma < \gamma_b$, the evolution of the eigenenergies shows energy level crossings as $\gamma$ increases, as seen in Fig.~\ref{fig2} (b), because of the orthogonality of the two separated energy bands. The eigenstates of the Hamiltonian are given by
\begin{equation}
\ket{\psi} = \alpha \ket{\psi_a} + \beta \ket{\psi_b},
\label{a_0}
\end{equation}
where $\ket{\psi_a} = (1,0)^{T}$ and $\ket{\psi_b} = (0,1)^{T}$. The corresponding eigenstates with unbroken phases are equally distributed to the upper and lower lattices, i.e., $\left|\alpha\right|^2 = \left|\beta\right|^2 = 0.5$, as shown in Fig.~\ref{fig3} (b). Using unitary matrix $U$, the Hamiltonian changes into a diagonalized Hamiltonian $H_\theta$ as,
\begin{equation}
U H U^{\dagger} = H_\theta,
\end{equation}
where
\begin{equation}
\label{rotation}
U =
\begin{pmatrix}
\cos{\theta} & -\sin{\theta} \\
\sin{\theta} & \cos{\theta}
\end{pmatrix}.
\end{equation}
From the diagonalized condition of $H_\theta$, the general rotating angle $\theta$ is derived as
\begin{equation}
\label{solution}
\theta = \frac{1}{2} \arccot{\left(\frac{i \gamma}{2 d}\right)}.
\end{equation}
Rotation angle $\theta$ is complex in a PT-symmetric CLL, while real in a Hermitian CLL with antisymmetric real on-site potential. Writing complex angle $\theta=\theta_r+i\theta_i$ for a general non-Hermitian system, the states can be transformed through the basis transformation with 
\begin{eqnarray}
\nonumber \frac{\delta+i\gamma}{2d}&=&\cot{2\theta}\\
&=&\frac{\cos{2\theta_r}\sin{2\theta_r}-i\sinh{2\theta_i}\cosh{2\theta_i}}{\sin^2{2\theta_r}\cosh^2{2\theta_i}+\cos^2{2\theta_r}\sinh^2{2\theta_i}},
\end{eqnarray}
which has two conditions from real and imaginary equations. For the Hermitian case, the rotating angle can be derived from this result in cases of $\theta_i=0$.
Here, PT symmetry gives a constraint in that the real part of the equation is zero and there are two regimes, PT-unbroken and broken phases, regardless of the fact that we only have an interest in the unbroken phases because of energy level crossings and PT-symmetric transitions.
In the region of bulk states with unbroken phases when $\gamma <\gamma_b$ in a PT-symmetric CLL, the transformation is satisfied by $\frac{\gamma}{2d}=\tanh(2\theta_i)$ with $\theta_r=\frac{\pi}{4}+n\frac{\pi}{2}~(n\in Z)$. 

Using the rotation of the basis states, the eigenstates $\ket{\psi_\theta}$ are given by
\begin{equation}
\ket{\psi_\theta} = \alpha_\theta \ket{\psi_a^{\theta}} + \beta_\theta \ket{\psi_b^{\theta}},
\label{a_th}
\end{equation}
where
\begin{eqnarray}
\ket {\psi_a^{\theta}} = U \ket{\psi_a}, \\
\ket {\psi_b^{\theta}} = U \ket{\psi_b}.
\end{eqnarray}
Applying rotation to the states, $|\alpha_{\theta}|^2$ and $|\beta_{\theta}|^2$ are changed into $0.0$ or $1.0$ as shown in Fig.~\ref{fig3} (d), while both $|\alpha|^2$ and $|\beta|^2$ are equal to $0.5$ as shown in Fig.~\ref{fig3} (b).
On the other hand, in the region of bulk broken phases when $\gamma > \gamma_b$, the transformation is satisfied by $\frac{\gamma}{2d}=\coth(2\theta_i)$ with $\theta_r=n\frac{\pi}{2}~(n\in Z)$.

Figure~\ref{fig2} (e--h) show evolutions of the eigenenergies in a PT-symmetric MLL with gain and loss on the upper and lower lattices, respectively, as $\gamma$ increases. The difference between the PT-symmetric CLL and the PT-symmetric MLL lies in the existence of PT phase transitions in the unbroken regime of the latter when $\gamma < \gamma_b$ and a pair of emergent interface states by complex energy inversion appears \cite{Ryu19}.
As shown in Fig.~\ref{fig2} (f), the PT phase transition in the PT-symmetric MLL takes place at a pair of EPs originating from degenerate points in the PT-symmetric CLL, which are diabolic points (not EPs) [Fig.~\ref{fig2} (b)].
The two EPs are developed with increasing PT-symmetric potential, where a pair of real energies merge at one EP and repel at the other, while a pair of imaginary energies repel at one EP and merge at the other. 
Through this process, the quantum states show PT phase transitions from unbroken to broken and back to unbroken states again. This effect is caused by the finite size and discrete lattice of the system in our MLL with an abrupt parameter change, of which the lengths between paired EPs of the non-orientability induced PT phase transitions decrease as the system size increases. We can clearly see the PT phase transitions through measurements of quantum transport in the lattice systems in the following section.

\begin{figure}
\begin{center}
\includegraphics[width=\figsizeone\textwidth]{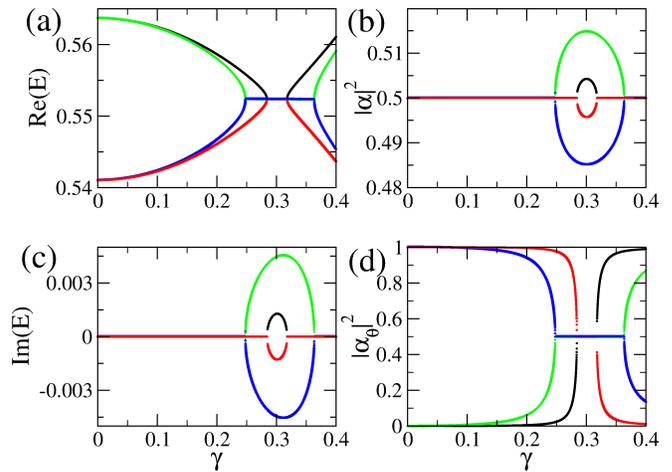}
\caption{(color online). Same as Fig.~\ref{fig3} but for the MLL. (a) Real and (c) imaginary parts of the eigenenergies in a PT-symmetric MLL. (b) $|\alpha|^2$ of four eigenstates and (d) $|\alpha_\theta|^2$ of four eigenstates.
The formation of a pair of EPs sandwiching a region of broken PT-phase ($\mathrm{Re}(E)=0$, $\mathrm{Im}(E) \ne 0$) is clearly visible.
}
\label{fig4}
\end{center}
\end{figure}

While there are many energy crossings in the unbroken region when $\gamma < \gamma_b$ in the case of a PT-symmetric CLL, many pairs of PT phase transitions appear in a PT-symmetric MLL. Figure~\ref{fig4} shows strong and weak PT phase transitions and also that a Hermitian degenerate point splits into a pair of non-Hermitian degenerate points, or EPs, due to the non-Hermiticity of this case. The strong or weak PT phase transition means that the region of broken states is wider or narrower. The eigenstates have unbroken phases, of which eigenenergies are real, before the PT phase transitions, where $|\alpha|^2=|\beta|^2=0.5$ as shown in Fig.~\ref{fig4} (b). In the region, $|\alpha_{\theta}|^2$ and $|\beta_{\theta}|^2$ decrease from $1.0$ or increase from $0.0$. After the PT phase transition, i.e., in the broken region, $|\alpha|^2$ and $|\beta|^2$ do not equal $0.5$ but are rather closely related to the imaginary parts of the eigenenergies. In the broken region, $|\alpha_{\theta}|^2$ and $|\beta_{\theta}|^2$ are equal to $0.5$. 

\section{Transport in corresponding ladder lattices}

\begin{figure}
\begin{center}
\includegraphics[width=\figsizeone\textwidth]{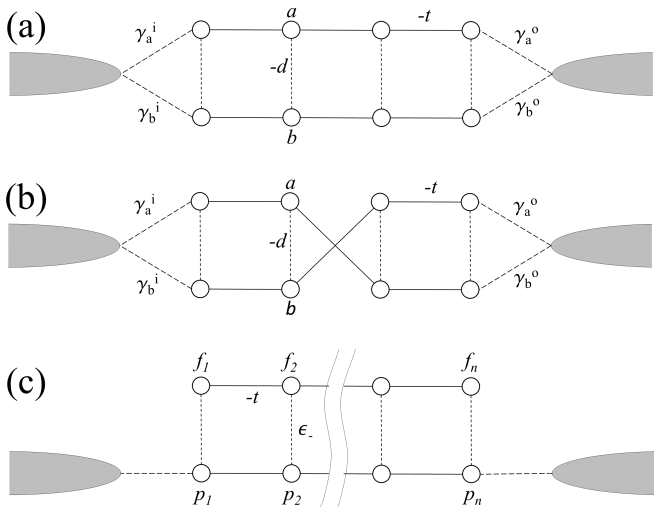}
\caption{(color online). (a) A ladder lattice with two leads. The unit cell has two sites, $a$ and $b$, with hopping strengths $d$ (dashed lines) and $t$ (solid lines) between the sites. The coupling strength between sites and leads is $\gamma_{u(d)}^{i(o)}$ (long-dashed lines). (b) A twisted-ladder lattice with two leads. (c) Detangled Fano lattices from a ladder lattice with symmetric contacts corresponding to (a). The hopping strength between $f_{n}$ and $p_{n}$ is $\epsilon_{-} = (\epsilon_{u} - \epsilon_{d})/2$.
}
\label{fig5}
\end{center}
\end{figure}

In this section, we discuss quantum transport in ladder and twisted-ladder lattices exhibiting energy-level crossings and PT phase transitions like as in the CLL and MLL. Figure~\ref{fig5} (a) and (b) illustrate the ladder and twisted-ladder lattices with two leads, respectively.

A reconfiguration of the quantum states and corresponding quantum transport in PT-symmetric lattices has previously been demonstrated in quasi-one-dimensional lattices \cite{Ryu17b, hcp}, where the quantum states can be controlled by PT-symmetric potential. In the broken phase, the group velocity has been found to be a complex number, of which the imaginary part is reflected in the suppression of the transmission due to the attenuation of the evanescent modes. As a result, quantum transport in the lattice with normal metallic leads is measured only in the unbroken PT-phases in the energy band (i.e., real spectra), not in the broken PT-phases.

Let us consider quantum transport in a ladder lattice with antisymmetric imaginary on-site potential, i.e., $\varepsilon_u=-\varepsilon_d= i \gamma/2$.
The system under study is composed of a ladder lattice with $N$ unit cells, as shown in Fig.~\ref{fig5} (a), with two leads connected to the left and right end unit cells.
The Hamiltonian of this system is given by
\begin{equation}
 H = H_{LL} + H_{lead} + H_{coupling},
\label{eq_trans}
\end{equation}
where $H_{LL}$, $H_{lead}$, and $H_{coupling}$ describe the ladder lattice, leads, and coupling between the lattice and leads, respectively, and are given by
\begin{eqnarray}
\label{eq_trans2}
 H_{LL} &=& \sum_{i=1}^{N}H_0 d_{i}^{\dagger} d_{i} + \sum_{i=1}^{N-1} ~(H_{1} d_{i+1}^{\dagger} d_{i} + h.c.) \\
 H_{lead} &=& -\frac{V_0}{2} \sum_{j \neq 0} (c_{j+1}^{\dagger} c_{j} + h.c.) \\
 H_{coupling} &=& - G^i d_{1}^{\dagger} c_{-1} - G^o d_{N}^{\dagger} c_{1} + h.c.,
\end{eqnarray}
where
\begin{eqnarray}
\label{h0h1}
H_0 = 
\begin{pmatrix} 
\epsilon_u  & -d \\
-d & \epsilon_d
\end{pmatrix} \text{,} \quad
H_1 =
\begin{pmatrix} 
-t  & 0 \\
0 & -t
\end{pmatrix},
\end{eqnarray}
and $d_{j}^{\dagger}$ ($d_j$) and $c_{j}^{\dagger}$ ($c_j$) are particle creation (annihilation) operators for the lattice and leads, respectively. $V_0 / 2$ is the hopping strength in the leads and $G^{i(o)}$ describes the coupling between the lattice and the left (right) lead. Considering a twisted-ladder lattice with two leads as in Fig.~\ref{fig5} (b), Eq.~(\ref{eq_trans2}) has to be changed into
\begin{eqnarray}
 \nonumber
 H_{tLL} &=& \sum_{i=1}^{N}H_0 d_{i}^{\dagger} d_{i} + \sum_{i\neq N/2, i=1}^{N-1} ~(H_{1} d_{i+1}^{\dagger} d_{i} + h.c.)  \\
 & & +  H_1{'} (d_{N/2+1}^{\dagger} d_{N/2} + h.c.),
\end{eqnarray}
where
\begin{eqnarray}
\label{h1prime}
H_1^{'} =
\begin{pmatrix} 
0  & -t \\
-t & 0
\end{pmatrix},
\end{eqnarray}
which induces cross coupling such as in the MLL.

\begin{figure*}
\begin{center}
\includegraphics[width=\figsizethree\textwidth]{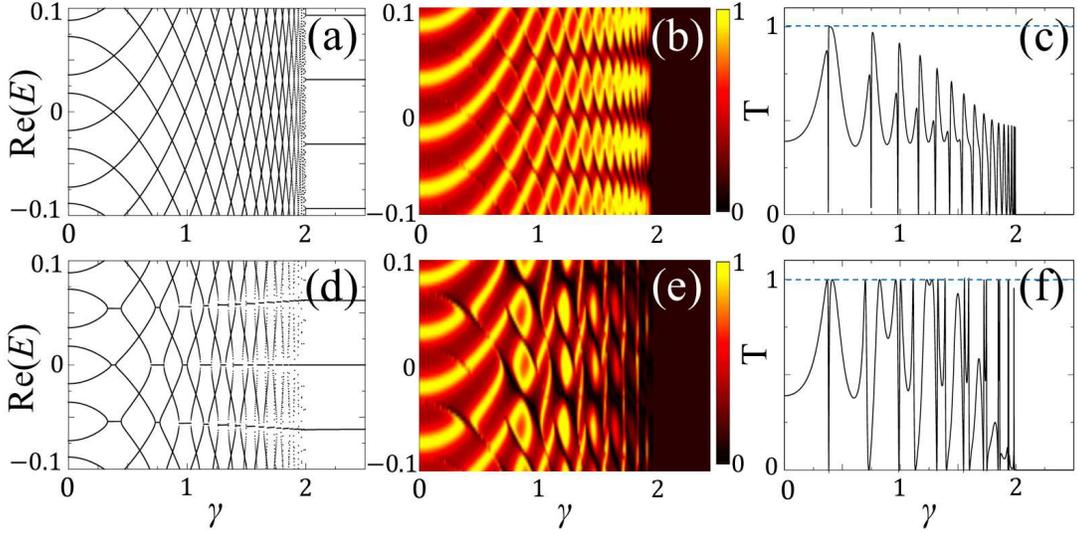}
\caption{(color online).
(a) Eigenenergies as a function of imaginary antisymmetric on-site potential $\gamma$ in a ladder lattice with $100$ unit cells. (b) Transmission probability for the ladder lattice in which both $a$ and $b$ sites of the end unit cells are connected to the input and output leads. Yellow, red, and black colors denote the highest, middle, and lowest transmission probabilities, respectively. (c) Transmission probability as a function of $\gamma$ when $\mathrm{Re}(E) = 0$. (d) Eigenenergies as a function of $\gamma$ in a ladder lattice with $100$ unit cells and one twisted hopping. (e) Transmission probability for the twisted-ladder lattice in which both $a$ and $b$ sites of the end unit cells are connected to the input and output leads. (f) Transmission probability as a function of $\gamma$ when $\mathrm{Re}(E) = 0$.
The dark regions in (e) appear larger than in (b) reflecting the presence of evanescent modes corresponding to imaginary eigenenergies in the broken PT-phase.
}
\label{fig6}
\end{center}
\end{figure*}

We construct ladder and twisted ladder lattices of $100$ unit cells with symmetric contacts---that is, $\gamma_u^{i} = \gamma_d^{i} = \gamma_u^{o} = \gamma_d^{o} = \gamma_0$, and $t = d = 1$. Throughout this work, we set $\gamma_{0}=1$. Figure ~\ref{fig6} (a) and (d) show the eigenenergy spectra of the ladder and twisted-ladder lattices, respectively. Using the above Hamiltonian, transmission probability $T=\left|\bm{t}\right|^2$ can be obtained as a function of $\gamma$ and energy $E$, as shown in Fig.~\ref{fig6} (b) and (e) corresponding to each lattice [see Appendix]. It is noted that $|\bm{r}|^2 + |\bm{t}|^2 = 1$ in PT-symmetric systems where $|\bm{r}|^2$ and $|\bm{t}|^2$ are the reflection and transmission probabilities, respectively, whereas $|\bm{r}|^2 + |\bm{t}|^2 \neq 1$ in non-Hermitian systems that do not preserve PT-symmetry. 

The paired eigenenergies of the ladder lattices are attracted under PT symmetry while the energies are repelled in the Hermitian system.
For the ladder lattice, the unbroken PT-phase shows level crossings of real eigenenergies and the broken PT-phase, $\gamma > \gamma_b$, maintains constant real eigenenergies as shown in Fig.~\ref{fig6} (a). For the twisted-ladder lattice, on the other hand, the energy spectra show PT phase transitions at multiple pairs of EPs corresponding to the energy level crossings in the unbroken region in the PT-symmetric ladder lattice, as shown in Fig.~\ref{fig6} (d).
It is from the features that distinguish the twisted-ladder from the ladder lattices that the level crossings and PT phase transitions are developed as a function of on-site potential $i \gamma$.

The level crossings and PT phase transitions are reflected by resonant and antiresonant peaks through Breight--Wigner and Fano resonances by means of quantum transport in both PT symmetric ladder lattices, as shown in Fig. \ref{fig6} (b,c) and (e,f). Transmission can be well explained by the energy spectra in both PT-symmetric ladder and twisted-ladder lattices; the transmission peaks trace the energy spectra of the systems.
The attached leads, however, perturb the system and modify the self-energy as a non-Hermitian parameter so that the PT symmetry of the system is broken. This modification lifts the degeneracy points, which is reflected by a transmission suppression at the points.
For the simple ladder lattice, the resonance peaks at zero energy are suppressed with increasing non-Hermitian parameter $\gamma$, even in the resonant condition. A leakage of transmissions is reflected but current flux is still conserved, similar to the Hermitian systems. 
For the twisted-ladder lattice, as aforementioned, the twisted boundary causing non-orientability introduces a PT phase transition at the crossed levels, and this PT phase transition can be captured through resonant transmission. The twisted boundary reveals perfect resonance, since the attracted energies coalesce at the EP characterized by self-orthogonality.
This finding, that non-orientability reveals perfect transmission at zero energy through the PT phase transition, is notable. 

In order to understand the antiresonant states, let us consider the amplitude equations for a ladder lattice. The equations of the $n$-th unit cell in a ladder lattice can be written as
\begin{equation}
E \bm{v_n}=\left(\frac{i \gamma}{2}\sigma_z-d\sigma_x\right)\bm{v_n}-t\sigma_0(\bm{v_{n-1}}+\bm{v_{n+1}}),
\label{eq_ll}
\end{equation}
where $\bm{v_n}=(a_n,b_n)$ is amplitude vector of the $n$-th unit cell. Applying Eq.~(\ref{rotation}) to these equations, we obtain amplitude equations of rotated states, $\bm{g_n}=(f_n,p_n)=U(\theta)\bm{v_n}$, as follows,
\begin{eqnarray}
\nonumber
E f_n &=& ((i \gamma /2)\cos{2\theta} + d \sin{2\theta}) f_n \\
 && - t (f_{n-1}+f_{n+1}) + \mathcal{F}(\theta) p_n, \\\nonumber
E p_n &=& (- (i \gamma /2)\cos{2\theta} - d \sin{2\theta}) p_n \\
 && - t (p_{n-1}+p_{n+1}) + \mathcal{F}(\theta) f_n,
\end{eqnarray}
where $\theta$ is real and $\mathcal{F}(\theta) = ((i \gamma /2)\sin{2\theta} - d \cos{2\theta})$ is a coupling term for the rotated equations that gives a detangled condition, Eq.~(\ref{solution}), when this term is zero. The condition provides that the states are completely separated into two orthogonal states, $f_n$ and $p_n$, while exhibiting energy crossings. Here, the orthogonal basis is associated with the symmetry of the leads, which is helpful for understanding the measurement process. When the rotation angle is $\theta = \pi/4$, we get
\begin{eqnarray}
E \bm{g_n} = \left(-d \sigma_z + \frac{i \gamma}{2} \sigma_x \right) \bm{g_n} - t \sigma_0(\bm{g_{n-1}}+\bm{g_{n+1}}),
\label{eq_det}
\end{eqnarray}
where the components of $\bm{g_n}$, $f_n = ( a_n - b_n ) / \sqrt{2}$, and $p_n = ( a_n + b_n ) / \sqrt{2}$, are antisymmetric and symmetric configurations, respectively.  For the symmetric contacts with leads as shown in Fig.~\ref{fig5} (c), $p_n$ and $f_n$ states result in resonance and antiresonance in transmission probability as a function of incoming energy and on-site potential, respectively, in Fig.~\ref{fig6}. If we use an asymmetric contact, e.g., $\gamma_u^{i} = \gamma_u^{o} = 0$, and $\gamma_d^{i} = \gamma_d^{o} = \gamma_0$, both lattices with on-site potentials $f_n$ and $p_n$ result in resonances, with no antiresonance.

There is a crucial difference between Hermitian and non-Hermitian PT-symmetric systems, namely whether the coupling between $f_n$ and $g_n$ in Eq.~(\ref{eq_det}) is real or imaginary according to the system possessing real mass or PT-symmetric mass.
This imaginary coupling drives the attraction between the real parts of the paired energy bands and then shows a PT phase transition. The transition is at a collective EP in both PT-symmetric lattices at $\gamma=\gamma_b$ in Fig.~\ref{fig6} (a) and (d).
While the transmission probability shows resonant features of the bulk states in unbroken PT-phases, $\gamma < \gamma_b$, the transmission probability is completely suppressed in the broken phase, $\gamma > \gamma_b$, due to the lack of resonant energy in the real energy space, as shown in Fig.~\ref{fig6} (b) and (e). The real eigenenergies of the unbroken PT phases correspond to propagating modes, while the complex eigenenergies of the broken PT phases are reflected in evanescent modes. As a result, in the broken PT phase, transmission is suppressed due to the attenuation of the evanescent modes irrespective of the details of the system. In contrast, the resonant transmission features seen in the unbroken regime reflect real eigenenergies.

\section{Summary}

We have studied the energy spectra of PT-symmetric ladder lattices containing non-orientability and corresponding quantum transport. Energy crossings and PT phase transitions in both circular and M{\"o}bius ladder lattices have been demonstrated and explained using generalized rotational transformation. Quantum transport in non-Hermitian PT-symmetric ladder lattices without and with a sharply twisted interface, corresponding to the circular and M{\"o}bius ladder lattices, respectively, have also been studied. Transmission probabilities show resonance and antiresonance in the energy spectra exhibiting energy crossings and PT phase transitions. Notably, the perfect transmission probability at the zero energy is recovered by the PT phase transition. We expect the combination of non-Hermiticity with real-space topological structures, like the PT-symmetric M{\"o}bius ladder lattice, to broaden the horizon of applications beyond existing non-Hermitian systems.

\section*{Acknowledgments}

This work was supported by Project Code (IBS-R024-D1), a National Research Foundation of Korea (NRF) grant (NRF-2019R1F1A1051215), and the Korea Institute for Advanced Study (KIAS) funded by the Korean government.

\section*{Appendix}

\subsection{Transmission probability}

The amplitude equations of the total Hamiltonian of Eq. (\ref{eq_trans}) can be written as
\begin{eqnarray}
 E \phi_{-1} &=& -\frac{V_0}{2} \phi_{-2} + {G^i}^T  \Psi_1\\
 \label{leftc}
 E \Psi_{1} &=& H_{0} \Psi_{1} + H_{1} \Psi_{2} + \phi_{-1} {G^i} \\
 E \Psi_{j} &=& H_{0} \Psi_{j} + H_{1}^{\dagger} \Psi_{j-1} + H_{1} \Psi_{j+1} (2 \leq j \leq N-1) \\
 \label{rightc}
 E \Psi_{N} &=& H_{0} \Psi_{N} + H_{1}^{\dagger} \Psi_{N-1} + \phi_{1} {G^o}\\
 E \phi_{1} &=& -\frac{V_0}{2} \phi_{2} + {G^o}^T \Psi_N 
\end{eqnarray}
where
\begin{eqnarray}
 \phi_{j} =& e^{i q j} + r e^{-i q j} & (j < 0) \\
          =& t e^{i q j} & (j > 0). 
\end{eqnarray}
Here, $\phi_{j}$ represents the $j$th sites of the leads and $G^{i(o)}$ is given by
\begin{equation}
\label{geq}
G^{i(o)} = \left(\begin{array}{c}
 -\gamma_u^{i(o)} \\
 -\gamma_d^{i(o)} 
\end{array}\right),
\end{equation}
and $r$ and $t$ are reflection and transmission coefficients, respectively, with $\left|r\right|^2 + \left|t\right|^2 = 1$ in the Hermitian case.
We then obtain the following equations:
\begin{eqnarray}
 -\frac{V_0}{2} &=& \frac{V_0}{2} r + {G^i}^T \Psi_1 \\
 - e^{-i q} {G^i} &=& e^{i q} r {G^i} + (H_{0} - E) \Psi_{1} + H_{1} \Psi_{2} \\
 0 &=& H_{1}^{\dagger} \Psi_{j-1} + (H_{0} - E) \Psi_{j} + H_{1} \Psi_{j+1} \\
 0 &=& H_{1}^{\dagger} \Psi_{N-1} + (H_{0} - E) \Psi_{N} + e^{i q} t {G^o} \\
 0 &=& \frac{V_0}{2} t + {G^o}^T \Psi_N,
\end{eqnarray}
where the energy of the leads is given by $e^{\pm i q} = - E / V_{0} \pm i \sqrt{1 - \left|E / V_{0}\right|^2}$.
Finally, we can obtain $R$ and $T$ for the ladder lattice from the following equation:
\begin{widetext}
\begin{equation}
 \left(\begin{array}{c}
 -\frac{V_0}{2} \\
 - e^{-iq} G^{i} \\
 0 \\
 \vdots \\
 0 \\
 0 \\
 0 \\
\end{array}\right)=
 \left(\begin{array}{ccccccc}
 \frac{V_0}{2} & {G^{i}}^T & & & & & \\
  e^{iq} G^{i} & H_0 - E I & H_1 & & & &  \\
  & H_1^{\dagger} & H_0 - E I & H_1 & & &  \\
  & & \ddots & \ddots & \ddots && \\
  &&& H_1^{\dagger} & H_0 - E I & H_1 & \\
  &&&& H_1^{\dagger} & H_0 - E I & e^{iq} G^{o}\\
  &&&&& {G^{o}}^T & \frac{V_0}{2} \\
\end{array}\right)
\left(\begin{array}{c}
 r \\
 \Psi_1 \\
 \Psi_2 \\
 \vdots \\
 \Psi_{N-1} \\
 \Psi_{N} \\
 t \\
\end{array}\right).
\end{equation}
\end{widetext}
Hamiltonians $H_0$ and $H_1$ are $2 \times 2$ matrices that describe the unit cell and the coupling between nearest unit cells, respectively. 
We set $V_0 = 10$ throughout this paper.


\begin{thebibliography}{150}

\bibitem{Moi11} N. Moiseyev, {\it Non-Hermitian Quantum Mechanics}, (Cambridge University Press, New York, 2011).

\bibitem{Guo09} A. Guo, G. J. Salamo, D. Duchesne, R. Morandotti, M. Volatier-Ravat, V. Aimez, G. A. Siviloglou, and D. N. Christodoulides, Observation of PT-Symmetry Breaking in Complex Optical Potentials, Phys. Rev. Lett. {\bf 103}, 093902 (2009).
\bibitem{Rue10} C. E. R{\"u}ter, K. G. Makris, R. El-Ganainy, D. N. Christodoulides, M. Segev, and D. Kip, Observation of parity-time symmetry in optics, Nat. Phys. {\bf 6}, 192-195 (2010).

\bibitem{Lie12} M. Liertzer, L. Ge, A. Cerjan, A. D. Stone, H. E. T{\"u}reci, and S. Rotter, Pump-induced exceptional points in lasers. Phys. Rev. Lett. {\bf 108}, 173901 (2012).
\bibitem{Bra14} M. Brandstetter, M. Liertzer, C. Deutsch, P. Klang, J. Sch{\"o}berl, H. E. T{\"u}reci, G. Strasser, K. Unterrainer, and S. Rotter, Reversing the pump dependence of a laser at an exceptional point. Nat. Commun. {\bf 5}, 4034 (2014).

\bibitem{Hod17} H. Hodaei, A. U. Hassan, S. Wittek, H. Garcia-Gracia, R. El-Ganainy, D. N. Christodoulides, and M. Khajavikhan, Enhanced sensitivity at higher-order exceptional points. Nature {\bf 548}, 187-191 (2017).
\bibitem{Che17} W. Chen, S. Kaya {\"O}zdemir, G. Zhao, J. Wiersig, and L. Yang, Exceptional points enhance sensing in an optical microcavity. Nature {\bf 548}, 192-196 (2017).

\bibitem{Moe65} A. F. M{\"o}bius, {\"U}ber die Bestimmung des Inhaltes eines Polyeders (On the determination of the volume of a polyhedron), Ber. Verh. S{\"a}chs. Ges. Wiss. {\bf 17}, 31-68 (1865); {\it Gesammelte Werke, Band II (Collected Works, vol. II)}, Hirzel, Leipzig (1886).

\bibitem{Lis61} J. B. Listing, der Census r{\"a}umlicher Complexe, oder Verallgemeinerung des Euler'schen Satzes von den Poly{\"a}dern, Abhandlungen der K{\"o}niglichen Gesellschaft der Wissenschaften in G{\"o}ttingen {\bf 10}, 97-182 (1862).

\bibitem{Spi65} M. Spivak, {\it Calculus on Manifolds: A Modern Approach to Classical Theorems of Advanced Calculus}, (HarperCollins, 1965).
\bibitem{Hat01} A. Hatcher, {\it Algebraic Topology}, (Cambridge University Press, 2001).

\bibitem{Pic06} C. A. Pickover, {\it The M{\"o}bius Strip: Dr. August M{\"o}bius's Marvelous Band in Mathematics, Games, Literature, Art, Technology, and Cosmology}, (Thunder's Mouth Press, 2006).

\bibitem{Wal82} D. M. Walba, R. M. Richards, and R. C. Haltiwanger, Total Synthesis of the First Molecular M{\"o}bius Strip, J. Am. Chem. Soc. {\bf 104}, 3219-3221 (1982).
\bibitem{Gil15} G. Gil-Ram{\'i}rez, D. A. Leigh, and A. J. Stephens, Catenanes: Fifty Years of Molecular Links, Angew Chem Int Ed., {\bf 54}, 6110-6150 (2015).
\bibitem{Han10} D. Han, S. Pal, Y. Liu, and H. Yan, Folding and cutting DNA into reconfigurable topological nanostructures, Nature Nanotechnology {\bf 5}, 712-717 (2010).

\bibitem{Kat96} T. Kato, {\it Perturbation Theory of Linear Operators}, (Springer, Berlin, 1996).
\bibitem{Hei12} W. D. Heiss, The physics of exceptional points. J. Phys. A {\bf 45}, 444016 (2012).

\bibitem{Dem01} C. Dembowski, H.-D. Gr{\"a}f, H. L. Harney, A. Heine, W. D. Heiss, H. Rehfeld, and A. Richter, Experimental Observation of the Topological Structure of Exceptional Points, Phys. Rev. Lett. {\bf 86}, 787 (2001).
\bibitem{Dem03} C. Dembowski, B. Dietz, H.-D. Gr{\"a}f, H. L. Harney, A. Heine, W. D. Heiss, and A. Richter, Observation of a Chiral State in a Microwave Cavity, Phys. Rev. Lett. {\bf 90}, 034101 (2003).
\bibitem{Lee09} S.-B. Lee, J. Yang, S. Moon, S.-Y. Lee, J.-B. Shim, S. W. Kim, J.-H. Lee, and K. An, Observation of an Exceptional Point in a Chaotic Optical Microcavity, Phys. Rev. Lett. {\bf 103}, 134101 (2009).

\bibitem{Gao15} T. Gao, E. Estrecho, K. Y. Bliokh, T. C. H. Liew, M. D. Fraser, S. Brodbeck, M. Kamp, C. Schneider, S. H{\"o}fling, Y. Yamamoto, F. Nori, Y. S. Kivshar, A. G. Truscott, R. G. Dall, and E. A. Ostrovskaya, Observation of non-Hermitian degeneracies in a chaotic exciton-polariton billiard, Nature {\bf 526}, 554-558 (2015).

\bibitem{Ben98} C. M. Bender and S. Boettcher, Real Spectra in Non-Hermitian Hamiltonians Having PT Symmetry, Phys. Rev. Lett. {\bf 80}, 5243 (1998).

\bibitem{Elg07} R. El-Ganainy, K. G. Makris, D. N. Christodoulides, and Z. H. Musslimani, Theory of coupled optical PT-symmetric structures, Opt. Lett. {\bf 32}, 2632 (2007).
\bibitem{Mus08} Z. H. Musslimani, K. G. Makris, R. El-Ganainy, and D. N. Christodoulides, Optical Solitons in PT Periodic Potentials, Phys. Rev. Lett. {\bf 100}, 030402 (2008).
\bibitem{Mak08} K. G. Makris, R. El-Ganainy, D. N. Christodoulides, and Z. H. Musslimani, Beam Dynamics in PT Symmetric Optical Lattices, Phys. Rev. Lett. {\bf 100}, 103904 (2008).
\bibitem{Kla08} S. Klaiman, U. G{\"u}nther, and N. Moiseyev, Visualization of Branch Points in PT-Symmetric Waveguides, Phys. Rev. Lett. {\bf 101}, 080402 (2008).

\bibitem{Reg12} A. Regensburger, C. Bersch, M. A. Miri, G. Onishchukov, D. N. Christodoulides, and U. Peschel, Parity-time synthetic photonic lattices, Nature (London) {\bf 488}, 167-171 (2012).
\bibitem{Sch11} J. Schindler, A. Li, M. C. Zheng, F. M. Ellis, and T. Kottos, Experimental study of active LRC circuits with PT symmetries, Phys. Rev. A {\bf 84}, 040101 (2011).
\bibitem{Jog10} Y. N. Joglekar, D. Scott, M. Babbey, and A. Saxena, Robust and fragile PT-symmetric phases in a tight-binding chain, Phys. Rev. A {\bf 82}, 030103(R) (2010).
\bibitem{Laz13} N. Lazarides and G. P. Tsironis, Phys. Gain-Driven Discrete Breathers in PT-Symmetric Nonlinear Metamaterials, Rev. Lett. {\bf 110}, 053901 (2013).

\bibitem{Zha09} N. Zhao, H. Dong, S. Yang, and C. P. Sun, Observable topological effects in molecular devices with M{\"o}bius topology, Phys. Rev. B {\bf 79}, 125440 (2009).

\bibitem{Guo09b} Z. L. Guo, Z. R. Gong, H. Dong, and C. P. Sun, M{\"o}bius graphene strip as a topological insulator, Phys. Rev. B {\bf 80}, 195310 (2009).
\bibitem{Kre18} J. Kreismann and M. Hentschel, The optical M{\"o}bius strip cavity: Tailoring geometric phases and far fields, EPL {\bf 121}, 24001 (2018). 

\bibitem{Guy67} R. K. Guy and F. Harary, On the M{\"o}bius ladders, Canadian Mathematical Bulletin, {\bf 10}, 493-496 (1967).

\bibitem{zli12} Z. Li and L. R. Ram-Mohan, Phys. Rev. B {\bf 85}, 195438 (2012).

\bibitem{Ryu17a} J.-W. Ryu, N. Myoung, and H. C. Park, Antiresonance induced by symmetry-broken contacts in quasi-one-dimensional lattices, Phys. Rev. B {\bf 96}, 125421 (2017).

\bibitem{Ryu19} J.-W. Ryu et al., Emergent localized states at the interface of a twofold PT-symmetric lattice, Phys. Rev. Research, {\bf 2}, 033149 (2020).

\bibitem{Ryu17b} J.-W. Ryu, N. Myoung, and H. C. Park, Reconfiguration of quantum states in PT-symmetric quasi-one-dimensional lattices, Scientific reports {\bf 7}, 8746 (2017).

\bibitem{hcp} H. C. Park, N. Myoung, J.-W. Ryu, Quantum Transport and Non-Hermiticity on Flat-Band
Lattices, J Low Temp. Phys., {\bf 191}, 49 (2018).

\end{thebibliography}
\end{document}